# The Y Dwarf Population with HST: unlocking the secrets of our coolest neighbours – II. Parallaxes and Proper Motions[⋆]

Clémence Fontanive,[1]† Luigi R. Bedin,[2] Loïc Albert,[1] and Daniella C. Bardalez Gagliuffi[3,4]

[1] *Trottier Institute for Research on Exoplanets, Université de Montréal, Montréal, H3C 3J7, Québec, Canada*
[2] *INAF-Osservatorio Astronomico di Padova, Vicolo dell'Osservatorio 5, I-35122 Padova, Italy*
[3] *Department of Physics & Astronomy, Amherst College, 25 East Drive, Amherst, MA 01003, USA*
[4] *Astrophysics Department, American Museum of Natural History, 79th Street at Central Park West, New York, NY 10024, USA*

**ABSTRACT**
We present astrometric results from a *Hubble Space Telescope* (*HST*) campaign aimed at determining precise distances for cold Y-type brown dwarfs. Combining observations from a dedicated *HST*/WFC3 program with archival data, we derive astrometric solutions for 15 nearby Y dwarfs, by linking the high-precision relative astrometry from *Hubble* to the high-accuracy *Gaia* DR3 absolute reference system, using stars present in both to anchor the two frames of reference. We reach uncertainties on parallaxes below the 1-mas level for half of the sample, and down to 3 mas for two thirds of the targets, or relative precisions <1% in most cases and 2–5× improvements over previous measurements. For the remaining targets, we achieved slightly lower precisions on parallaxes (5–12 mas, 5–10%), correlated with the lower signal-to-noise of the faintest targets. The precision reached in our derived proper motions is around 0.1–0.4 mas/yr for most targets, and up to 1–2 mas/yr for less precise cases. Our estimated parallaxes and proper motions are generally in good agreement with literature values, and consistent to 1–2$\sigma$ with recent *Spitzer*-derived parallaxes in most cases. These new astrometric solutions provide important validation of these objects' distances and sky motions, especially given the large disparities seen in previous estimates. Our results demonstrate the power of *HST* combined with *Gaia* to measure highly-precise absolute astrometry for faint brown dwarfs, and highlights the limitations reached for the reddest and coldest objects, for which *JWST* will certainly provide a favourable platform to improve these results.

**Key words:** brown dwarfs – astrometry – parallaxes – proper motions

## 1 INTRODUCTION

With effective temperatures below 500 K, Y dwarfs (Cushing et al. 2011) bridge the warmer populations of L- and T-type brown dwarfs and young giant exoplanets, to cooler Solar System gas giants like Jupiter, and hold the key to fundamental astrophysical questions, including the universality and limits of the initial mass function (IMF; the mass distribution of stars and brown dwarfs), one of the biggest open questions in astronomy. With estimated masses of ∼3–25 $M_{\rm Jup}$ (Kirkpatrick et al. 2021), the current sample of free-floating Y dwarfs in the Solar neighbourhood represents the low-mass tail of the observed IMF. However, the current bottom end of the IMF likely remains incomplete, and its true end point is still unknown. Isolated Y dwarfs also provide state-of-the-art tests of cold Jupiter-like atmospheres, but while tremendous progress has been made in the atmospheric modelling of the coldest substellar objects (Phillips et al. 2020; Marley et al. 2021; Lacy & Burrows 2023), current theoretical predictions still struggle to replicate existing observations (e.g., Leggett et al. 2021; Fontanive et al. 2023).

The knowledge of distances to astronomical objects is essential for robust flux calibrations of individual systems, as well as to define volume-limited samples for statistical population studies. Measuring precise distances for free-floating Y dwarfs is thus fundamental to enable in-depth investigations of these cold planetary environments and test the extreme outcome of star-forming processes. The *James Webb Space Telescope* (*JWST*) infrared (IR) capabilities now provide an outstanding platform to observe in detail the thermal emission of cold and red Y dwarfs beyond 3–4 $\mu$m (e.g., Beiler et al. 2024), making the imperative for accurate distance calibrations even more pronounced.

While the *Gaia* mission has been successfully delivering reliable parallaxes for almost 2 billion stars, cold Y dwarfs are too dim and too red to be detected by *Gaia* at visible wavelengths. From the dedicated ground- and space-based campaigns that have attempted to measure the astrometry of Y dwarfs, the challenges associated with observations of such faint objects have resulted in large uncertainties and significant disparities between the measured parallaxes (e.g., Dupuy & Kraus 2013; Marsh et al. 2013; Beichman et al. 2014; Pinfield et al. 2014; Martin et al. 2018; Tinney et al. 2018). More recently, the results from a large *Spitzer* campaign by Kirkpatrick et al. (2019, 2021) provided homogeneous astrometry for the full Y dwarf population, down to ∼2-mas level on parallaxes, a significant improvement over previous measurements. However, the large

---

[⋆] Based on observations with the NASA/ESA *Hubble Space Telescope*, obtained at the Space Telescope Science Institute, which is operated by AURA, Inc., under NASA contract NAS 5-26555.
† E-mail: clemence.fontanive@umontreal.ca

discrepancies seen within the various existing distance estimates for most currently-known Y dwarfs motivates the need to confirm these with independent measurements.

In Bedin & Fontanive (2018, 2020), we devised an astrometric method linking multi-epoch *Hubble Space Telescope* (*HST*) observations to the *Gaia* catalogues, providing a powerful way to derive precise distances for dim sources invisible to *Gaia* but detected in near-IR *HST* images. In Fontanive et al. (2021), we introduced the first results from a dedicated *HST* program (GO 16229; PI Fontanive) awarded to apply the methodologies developed in Bedin & Fontanive (2018, 2020) to nearby Y dwarfs. Our test target validated the design of our program, reaching a remarkable sub-milliarcsecond precision in the derived parallax and proper motion with strategically-timed observations optimised for astrometry.

In this paper, we present astrometric results from this *HST* campaign for 15 Y dwarfs. Section 2 provides an overview of the *HST* program and archival observations used here. The data reduction and astrometric analyses are described in Section 3. The results are presented in Section 4, where we also discuss the performance and limitations of the procedures. Our conclusions are presented in Section 5.

## 2 HST DATA

### 2.1 GO 16229 Program Overview

Our *HST* program (GO 16229, PI Fontanive) was designed to measure some of the most precise parallaxes and proper motions to date for the majority of the ultracool Y dwarf population known at the time of the proposal submission. The observed sample consists of all *single* Y dwarfs confirmed at the time (based on the Y dwarf Compendium[1]) that had at least one archival epoch of imaging observations with the *HST* Wide Field Camera 3 (WFC3). WISE 0855−0714 (Luhman 2014), the coldest rogue Y dwarf identified to date (∼285 K; Luhman et al. 2024) was removed from the sample selection due to its intrinsic faintness in the near-IR, requiring deep multi-orbit exposures for robust detections in *HST* filters (e.g., Luhman & Esplin 2016; Schneider et al. 2016). We also excluded known Y dwarfs that are companions to more massive objects (Luhman et al. 2011; Liu et al. 2012; Dupuy et al. 2015). This provided us with a sample of 19 nearby Y-type brown dwarfs. Excluding the above systems and recent discoveries of new Y dwarfs (Marocco et al. 2019; Bardalez Gagliuffi et al. 2020; Meisner et al. 2020) from the CatWISE catalogue (Eisenhardt et al. 2020; Marocco et al. 2021), our campaign includes all but three of the confirmed Y dwarfs currently listed in Y dwarf Compendium, which did not satisfy the above criterion on existing archival epochs of *HST* data.

The program consists of 26 orbits of WFC3 IR channel imaging for precise photometry and astrometry of these selected ultracool Y dwarfs, aimed at complementing existing archival *HST* data of the targets. The primary goal of the program was to acquire the final observational epochs required to derive astrometric solutions (minimum of three epochs needed), with a secondary objective to obtain homogeneous sets of multi-band photometry in the F105W, F125W and F160W WFC3/IR filters on board *HST*. We focus this paper on the astrometric results and will present detailed analyses of the multi-band photometry in a subsequent paper.

Based on the number of past epochs and filters used in the archival material, we hence obtained the missing datasets needed to fulfil both the astrometric and photometric goals of the project. Specific time constraints were implemented for each visit in order to optimise the science results in the derivation of parallaxes and proper motions of each target. We defined optimal observing windows for each new orbit based on the yearly phases already covered by archival datasets and allowed visibility windows. When possible we aimed at acquiring a new epoch at a repeat of the time of year observed in a previous epoch and/or annual phases covering the maximum or minimum parallax elongations.

Here, we present astrometric results for 15 targets from the 19 in our observed sample. The targets studied in this paper are listed in Table 1, with their full *Wide-Field Infrared Survey Explorer* (*WISE*; Wright et al. 2010) designations, along with abbreviated source names to the short form WISE hhmm±ddmm used throughout the rest of this paper. Three targets from the observed sample have too few (≤3) reference *Gaia* stars in the fields of view for our current methodologies to be successfully implemented, and will be analysed separately in a following paper testing more advanced procedures. The fourth target, WISE 0336−0143, was recently discovered with *JWST*/NIRCam to be a tight binary, unresolved in existing *HST* images Calissendorff et al. (2023). As the astrometric approaches used here do not account for multiplicity, we will explore the *HST* astrometry of this binary system in a subsequent work, with additional data from a new *HST* program led by our team (GO 17466, PI Fontanive) that will allow to include the binary nature of the system in our astrometric analyses to fit its absolute motion and the binary orbit.

### 2.2 Details of Observations

The analyses presented in this paper are based on observations acquired in our GO 16229 *HST* program described above, but also relied and was designed based on archival data from past *HST* programs. The latest epochs always come from our own campaign, complementing previous *HST* observations and providing total time baselines of 7–10 years for all targets. All orbits in our program were planned as single-orbit visits, applying large and fractional-pixel dithering patterns to allow removal of hot pixels, improve the handling of the geometric distortion, and better sample the point spread functions (PSFs). All observations are taken in full frame mode, in order to maximise the number of reference sources in the fields of view. We used ∼300-s exposures in MULTIACCUM mode in all cases, varying the exact exposure times from NSAMP=12-13-14, with SAMP-SEQ=SPARS25, to fit within the orbital visibility of each orbit. This allowed for a total of 8 individual images per 1-orbit visit, crucial for precision astrometry within each epoch given the under-sampled detector of the WFC3 IR channel. The individual images were obtained in various combinations of the F105W, F125W and F160W filters depending on the photometry already available in archival data, with at least 2 exposures in a given band when using the F105W and F125W filters, and a minimum of 6 exposures when using the redder F160W band.

The archival observational epochs come from a range of *HST* programs that each had different goals and observing strategies. Some of these come from programs led by our team (GO 15201, PI Fontanive; GO 12873, PI Biller) for binary surveys of late-type brown dwarfs, and the F127M observations from these programs used here are described in details in Fontanive et al. (2018, 2023). Both programs also acquired observations in the F139M to validate the nature of identified companions by probing a water absorption band present in cool substellar objects around 1.4$\mu$m. However, none of our Y-dwarf primaries are recovered in the F139M images from these programs due to the depth of this feature at very late spectral types, and we

---

[1] https://sites.google.com/view/ydwarfcompendium/home

**Table 1.** Epochs of HST WFC3/IR observations used for the astrometric analyses presented in this work.

| Target Name | Short Name | SpT | Discovery Ref. | Program ID | Filter | $N_{images}$ | $t_{exp}$ (s) | Obs. Date |
|---|---|---|---|---|---|---|---|---|
| WISE J035000.32−565830.2 | WISE 0350−5658 | Y1 | Kirkpatrick et al. (2012) | GO-16229 | F160W | 8 | 2423 | 2021-03-27 |
| | | | | GO-16229 | F105W, F125W | 4, 4 | 1212, 1212 | 2021-01-05 |
| | | | | GO-15201 | F127M | 4 | 1312 | 2018-08-27 |
| | | | | GO-12330 | F140W | 4 | 412 | 2011-08-13 |
| WISEP J041022.71+150248.5 | WISE 0410+1502 | Y0 | Cushing et al. (2011) | GO-16229 | F160W | 8 | 2224 | 2021-02-28 |
| | | | | GO-16229 | F105W, F125W | 4, 4 | 1112, 1112 | 2020-09-11 |
| | | | | GO-15201 | F127M | 4 | 1199 | 2017-12-23 |
| | | | | GO-12544 | F140W | 4 | 312 | 2012-09-01 |
| WISE J053516.80−750024.9 | WISE 0535−7500 | ≥Y1 | Kirkpatrick et al. (2012) | GO-16229 | F160W | 8 | 2574 | 2021-05-23 |
| | | | | GO-16229 | F160W | $1^a$ | 278 | 2021-04-23 |
| | | | | GO-12970 | F105W | 3 | 434 | 2013-12-03 |
| | | | | GO-12970 | F105W, F125W | 6, 3 | 868, 808 | 2013-09-26/27 |
| | | | | GO-12544 | F140W | 4 | 312 | 2012-09-17 |
| | | | | GO-12330 | F140W | 4 | 512 | 2011-09-27 |
| WISE J064723.23−623235.5 | WISE 0647−6232 | Y1 | Kirkpatrick et al. (2013) | GO-16229 | F160W | $7^b$ | 2171 | 2021-05-11 |
| | | | | GO-12970 | F105W, F125W | 6, 3 | 818, 734 | 2013-12-27/30 |
| | | | | GO-12970 | F105W | 3 | 409 | 2013-11-14 |
| | | | | GO-12970 | F105W, F125W | 6, 3 | 818, 734 | 2013-05-13/14 |
| | | | | GO-12970 | F125W | 4 | 3015 | 2013-02-11 |
| WISE J071322.55−291751.9 | WISE 0713−2917 | Y0 | Kirkpatrick et al. (2012) | GO-16229 | F160W | 8 | 2224 | 2021-08-10 |
| | | | | GO-16229 | F160W | $6^c$ | 1668 | 2021-04-11 |
| | | | | GO-16229 | F105W, F125W | 4, 4 | 1112, 1112 | 2020-10-09 |
| | | | | GO-12873 | F127M | 2 | 698 | 2013-08-15 |
| WISE J073444.02−715744.0 | WISE 0734−7157 | Y0 | Kirkpatrick et al. (2012) | GO-16229 | F160W | 8 | 2574 | 2022-01-30 |
| | | | | GO-12873 | F127M | 2 | 698 | 2013-09-23 |
| | | | | GO-12970 | F105W, F125W | 4, 4 | 412, 412 | 2013-05-18/20 |
| | | | | GO-12970 | F125W | 4 | 2412 | 2013-02-04 |
| WISEA J082507.37+280548.2 | WISE 0825+2805 | Y0.5 | Schneider et al. (2015) | GO-16229 | F160W | 8 | 2224 | 2020-10-26 |
| | | | | GO-12970 | F105W, F125W | 9, 3 | 2127, 634 | 2014-01-16/19 |
| | | | | GO-12970 | F125W | 4 | 2412 | 2013-01-21 |
| WISEA J120604.25+840110.5 | WISE 1206+8401 | Y0 | Schneider et al. (2015) | GO-16229 | F160W | 8 | 2599 | 2021-01-24 |
| | | | | GO-15021 | F127M | 4 | 1312 | 2018-04-01 |
| | | | | GO-13178 | F105W, F125W | 4, 4 | 412, 412 | 2014-07-15 |
| | | | | GO-13178 | F125W | 4 | 2412 | 2013-03-18 |
| WISEP J154151.65−225025.2 | WISE 1541−2250 | Y1 | Cushing et al. (2011) | GO-16229 | F160W | $7^d$ | 1946 | 2021-08-14 |
| | | | | GO-15201 | F127M | 4 | 1197 | 2018-02-17 |
| | | | | GO-12970 | F105W, F125W | 4, 4 | 362, 312 | 2013-05-09 |
| | | | | GO-12970 | F125W | 4 | 2412 | 2013-02-12 |
| WISE J163940.83−684738.6 | WISE 1639−6847 | Y0pec | Tinney et al. (2012) | GO-16229 | F160W | 8 | 2524 | 2020-09-02 |
| | | | | GO-15201 | F127M | 4 | 1312 | 2019-03-11 |
| | | | | GO-12970 | F105W, F125W | 6, 4 | 768, 987 | 2013-10-26/30 |
| | | | | GO-12970 | F125W | 4 | 2412 | 2013-02-15 |
| WISEP J173835.52+273258.9 | WISE 1738+2732 | Y0 | Cushing et al. (2011) | GO-16229 | F105W, F125W | 4, 4 | 1112, 1112 | 2021-05-13 |
| | | | | GO-16229 | F160W | 8 | 2224 | 2020-09-09 |
| | | | | GO-15201 | F127W | 4 | 1200 | 2017-11-07 |
| | | | | GO-12330 | F140W | 4 | 312 | 2011-05-12 |
| WISEP J182831.08+265037.8 | WISE 1828+2650 | ≥Y2 | Cushing et al. (2011) | GO-16229 | F160W | 8 | 2224 | 2020-10-26 |
| | | | | GO-12970 | F105W, F125W | 3, 3 | 509, 559 | 2013-08-14/17 |
| | | | | GO-12970 | F105W | 6 | 1018 | 2013-06-21/25 |
| | | | | GO-12970 | F105W, F125W | 3, 3 | 509, 559 | 2013-05-06/08 |
| | | | | GO-12970 | F105W | 3 | 509 | 2013-04-22 |
| | | | | GO-12330 | F140W | 4 | 312 | 2011-05-09 |
| WISEPC J205628.90+145953.3 | WISE 2056+1459 | Y0 | Cushing et al. (2011) | GO-16229 | F160W | 8 | 2198 | 2021-05-01 |
| | | | | GO-16229 | F105W, F125W | 4, 4 | 1112, 1112 | 2020-11-17 |
| | | | | GO-15201 | F127M | 4 | 1197 | 2018-09-12 |
| | | | | GO-12330 | F140W | 4 | 312 | 2011-09-04 |
| WISE J220905.73+271143.9 | WISE 2209+2711 | Y0 | Cushing et al. (2014) | GO-16229 | F125W, F160W | 2, $5^e$ | 556, 1390 | 2020-11-02 |
| | | | | GO-12970 | F105W | 3 | 584 | 2013-09-20 |
| | | | | GO-12970 | F105W | 3 | 584 | 2013-06-06 |
| | | | | GO-12970 | F105W | 3 | 584 | 2013-04-28 |
| | | | | GO-12544 | F140W | 4 | 312 | 2012-09-15 |
| WISE J222055.31−362817.4 | WISE 2220−3628 | Y0 | Kirkpatrick et al. (2012) | GO-16229 | F160W | 8 | 2224 | 2021-12-10 |
| | | | | GO-12873 | F127M | 2 | 700 | 2013-07-21 |
| | | | | GO-12970 | F125W | 4 | 512 | 2013-06-20 |
| | | | | GO-12970 | F105W | 4 | 312 | 2013-06-08 |
| | | | | GO-12970 | F125W | 4 | 2412 | 2012-11-23 |

**Notes.**
Spectral types are the adopted spectral types for each source in the Y Dwarf Compendium.
$^a$ this epoch was failed and only image *ieay09zoq* was kept from the 8 individual exposures acquired.
$^b$ image *ieay10guq* was rejected due to its degraded quality.
$^c$ images *ieay12i7q* and *ieay12i9q* were rejected due to their degraded quality.
$^d$ image *ieay17bnq* was rejected due to its degraded quality.
$^e$ image *ieay24a8q* was rejected due to its degraded quality.

therefore do not consider these data here. The remaining observations come from GO 12330 (PI Kirkpatrick), GO 12544 (PI Cushing), GO 12970 (PI Cushing), and GO 13178 (PI Kirkpatrick), older *HST* programs aimed at confirming the nature of early *WISE* and *Spitzer* Y-dwarf candidates (Kirkpatrick et al. 2012; Cushing et al. 2014).

The observing programs, filters, number of individual images ($N_{images}$), and total exposure times ($t_{exp}$) for all $N_{images}$ exposures in each filter and epoch, are summarised in Table 1. Each line corresponds to a separate observational epoch for each target (grouped to windows within a few days), and multiple entries per line are listed when observations in multiple filters were acquired in a given epoch.

Some of the *HST* orbits acquired as part of our GO 16229 program failed due to delayed guide star acquisitions, causing measurements of stellar positions and fluxes to be compromised. In most cases, these orbits were successfully repeated at a later epoch. However, some exposures acquired during failed orbits were nonetheless found to be usable and are therefore considered here. Only individual images that were found to be too degraded in quality were rejected and excluded from the analyses performed in this paper. The numbers of frames given in Table 1 reflect the removal of these degraded images, which are listed in the Table notes.

## 3 ANALYSES

The strategies employed in this study for data reduction and analyses closely follow the protocols outlined in Bedin & Fontanive (2018, 2020) and Fontanive et al. (2021), updated from using the *Gaia* mission (Gaia Collaboration et al. 2016) Data Release 2 (Gaia Collaboration et al. 2018), to the now-available Data Release 3 (DR3) (Gaia Collaboration et al. 2023). We provide a brief summary of these methodologies here, but direct readers to the original works for comprehensive descriptions of the detailed procedures.

### 3.1 Data Reduction

For each target, the initial steps involved extracting positions and magnitudes of sources in each WFC3/IR flat-fielded image, using the publicly available software from Anderson & King (2006). A quality check was then applied to ensure the accuracy of measurements and filter out undesirable sources, such as galaxies or those affected by detector artifacts. Subsequently, raw pixel positions were corrected for the geometric distortion of the camera based on the WFC3/IR distortion solutions and PSFs derived by Anderson (2016)[2].

A reference frame corresponding to the latest *HST* epoch available for each target was then defined and linked to the *Gaia* DR3 absolute frame of reference (Gaia Collaboration et al. 2023), to transform our measured distortion-corrected positions into the ICRS reference frame. This was done by identifying all detected sources bright enough to be catalogued in *Gaia* DR3 and with full five-parameter astrometric solutions, and propagating their *Gaia* coordinates to the specific epochs of the various *HST* observations based on the *Gaia* parallaxes and proper motions, using the most general linear transformation. This procedure involves going back and forth between projected and tangential planes, using methods and equations detailed in section 3 of Bedin & Fontanive (2018) to link the coordinates within each individual *HST* image to the *Gaia* ICRS. The consistency among residuals between the reference and transformed *Gaia* positions were clipped at 3-sigma, with a dispersion of the order of 1 mas.

The same tasks were then performed for the remaining observational datasets, and the extracted measured positions in all the images from the different epochs were linked to the frame of the chosen reference epoch. This step established a common reference frame for each target, for all images across the various epochs available, defined by *Gaia* DR3 reference sources that can be re-positioned to the epoch of each individual image with their respective astrometric solutions. This approach enables the transformation of the positions of every source detected in the *HST* images, including fainter objects like our Y-dwarf targets that are not detectable by *Gaia*.

From these coordinate transformations into the common reference frame, we created stacked images for each available epoch in each filter, obtained as a clipped mean from a given epoch/filter combination, with prior removal of bad pixels based on the data quality (DQ) array information. The analyses presented in this paper are not based on these stacked images, but are instead performed entirely on the images before averaging. These stacked images provide a representation of the astronomical scene that can be used to independently check the nature of detected sources in each image (see Bedin & Fontanive 2018, 2020 for additional details about the stacked images). We make these astrometrised multi-band atlases publicly available[3], where world coordinate system keywords are included in the header of `fits` images. We also note that the coordinates discussed throughout this paper are not in the same pixel-coordinate system of these stacked images, which are instead supersampled by a factor of 2 (i.e., ∼60 mas in size).

### 3.2 Derivation of Astrometric Solutions

From the observed 2-D data points from each individual image, we then derived the five-parameter astrometric solution of each target, which include its position $(X, Y)$, proper motion $(\mu_X, \mu_Y)$ and parallax $(\varpi)$. We kept all measurements in the observational plane $(X, Y)$, as motivated in our method papers (Bedin & Fontanive 2018, 2020), and weighted each individual data point with the PSF quality-fit ($Q$; Anderson et al. 2008), that measures how well the flux distribution resembles the PSF models (close to 0 for well-behaved point sources, and approaching 1 or worse for compromised measurements). This allows to taken into account the fact that some measurements from individual exposures affected by blends, detector cosmetics or artefacts might be of poorer quality. Our primaries have average $Q$ values between 0.04–0.6 across all available images for each target, with a median for the full sample of $Q \sim 0.08$, and only a handful of targets with particularly low signal-to-noise detections having poorer quality-of-fit measurements (see Section 4.2).

We used the Naval Observatory Vector Astrometry Software (`NOVAS`) tool (Kaplan et al. 2011) to predict the position of the targets, which allows to account for effects like the Earth's orbit, perturbations of major bodies, nutation of the Moon-Earth system, etc. We then use a Levenberg-Marquardt algorithm (the FORTRAN version `lmdif` available under `MINIPACK`, Moré et al. 1980) to find the minimisation of the five parameters: $X, Y, \mu_X, \mu_Y$, and $\varpi$ and solve for the non-linear sky path solution.

As in our previous papers, given the nonlinearity in the $(X, Y)$ to $(\alpha, \delta)$ transformation, we assessed the uncertainties on our astrometric solutions with a Monte Carlo approach. For each target, we simulated 25 000 expected positions from the best-fit solution, with random errors following Gaussian distributions with bias-corrected standard deviations derived from the positional scatter in the observed data for each filter/epoch combination. The bias correction is used to provide an unbiased estimator of the variance and down-weight epochs with few points. For the one case in which there was only one useable image available (from a failed visit for WISE 0535−7500, see Table 1), we used the standard deviation from the repeated successful visit, acquired in the same filter and conditions. We then adopted the standard deviations of the resulting best-fit solutions to each of the Monte Carlo draws as 1-$\sigma$ uncertainties on the fitter parameters. No weighting from the measured $Q$ values was applied here, and we only applied a clipping to measurements with $Q > 1$ (9 images out of the total 376 considered here, around 5 different targets) before estimating the positional scatter within each dataset. This leaves us on the conservative side when estimating the uncertainties on the final astrometric solutions based on observed scatters, as this ap-proach retains more spurious and offset measurements that drive the considered variance up, despite having lower weights in the fits.

---

[2] the material is publicly available at https://www.stsci.edu/~jayander/WFC3/

[3] the stacked images are available at https://github.com/cfontanive/BDs_HST.

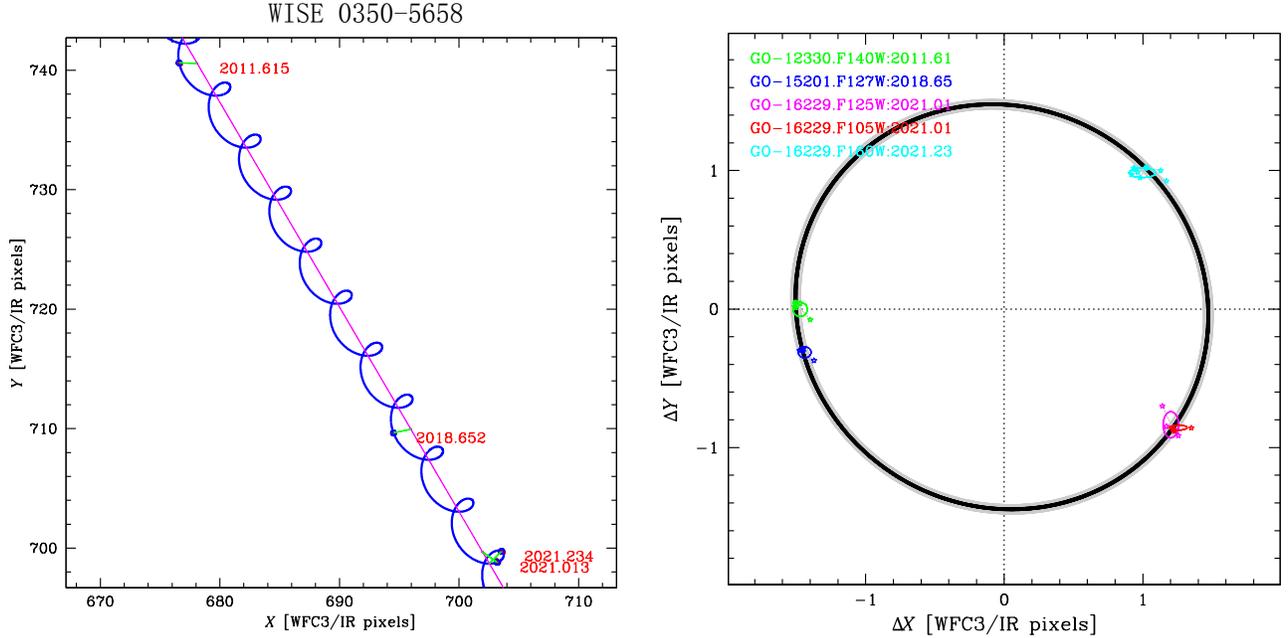

**Figure 1.** Derived astrometric solution for WISE 0350−5658. *Left*: Comparison of our astrometric solution (blue line) with the individual observed data points (red bullets) in the distortion-corrected observational plane at the reference epoch. The four major epochs are labelled. The sizes of the red circles around each bullet (barely visible in this case) indicate the quality-fit parameter $Q$ (Anderson et al. 2008) for each data point, with smaller radii for better measurements. To better highlight the parallax component of the motion, a magenta line marks the motion of an object with the same proper motion but placed at infinite distance (i.e., with zero parallax), with green lines showing the parallax contributions at each epoch according to the best fit. *Right*: Our solution for the parallax ellipse in the distortion-corrected reference coordinate system at the reference epoch, from the proper motion subtracted solution. Individual *HST* data points are indicated with star symbols, and their expected positions along the best-fit solution (black line) are shown with scatter points of the same colour. Smaller colour-coded ellipses indicate the standard deviation of the individual positional measurements in $X$ and $Y$ within each epoch and filter, and the final ±1-$\sigma$ uncertainty on the parallax is shown with grey lines.

**Table 2.** Derived absolute astrometric parameters of our targets in the ICRS reference frame. Positions are given at the Gaia DR3 2016.0 epoch.

| Target | $\alpha_{2016.0}$ [deg(mas)] | $\delta_{2016.0}$ [deg(mas)] | $\mu_\alpha \cos\delta$ [mas yr$^{-1}$] | $\mu_\delta$ [mas yr$^{-1}$] | $\varpi$ [mas] | $\chi^2_\nu$ | $\chi^2_{\nu,Q}$ |
|---|---|---|---|---|---|---|---|
| WISE 0350−5658 | 57.5007871(4.9) | −56.97595125(3.2) | −208.27 ± 0.71 | −572.75 ± 0.71 | 182.86 ± 3.07 | 1.41 | 0.88 |
| WISE 0410+1502 | 62.5964034(0.6) | +15.04318382(0.7) | +959.87 ± 0.17 | −2218.82 ± 0.18 | 153.01 ± 0.70 | 1.28 | 1.08 |
| WISE 0535−7500 | 83.8192173(10.8) | −75.00672406(2.9) | −119.88 ± 1.24 | +24.93 ± 1.29 | 77.49 ± 9.13 | 1.85 | 1.65 |
| WISE 0647−6232 | 101.8468007(6.6) | −62.54261151(2.2) | +1.19 ± 1.18 | +393.81 ± 0.78 | 110.06 ± 2.25 | 1.59 | 1.37 |
| WISE 0713−2917 | 108.3446413(2.1) | −29.29841594(1.7) | +352.26 ± 0.35 | −408.52 ± 0.32 | 110.11 ± 0.61 | 1.69 | 1.51 |
| WISE 0734−7157 | 113.6805030(1.8) | −71.96236286(1.1) | −565.67 ± 0.16 | −68.88 ± 0.16 | 73.35 ± 0.71 | 1.12 | 0.83 |
| WISE 0825+2805 | 126.2805153(4.4) | +28.09641437(2.5) | −68.76 ± 1.78 | −236.18 ± 0.94 | 160.22 ± 10.25 | 1.44 | 1.12 |
| WISE 1206+8401 | 181.5094424(13.1) | +84.01913953(4.7) | −576.17 ± 0.47 | −261.55 ± 0.34 | 83.92 ± 4.96 | 1.03 | 0.70 |
| WISE 1541−2250 | 235.4635169(1.6) | −22.84059716(1.0) | −903.02 ± 0.41 | −87.93 ± 0.27 | 172.28 ± 1.55 | 1.17 | 0.82 |
| WISE 1639−6847 | 249.9226303(1.7) | −68.79900761(0.5) | +577.09 ± 0.24 | −3108.07 ± 0.13 | 212.67 ± 0.91 | 1.22 | 0.98 |
| WISE 1738+2732 | 264.6486526(1.1) | +27.54912533(1.0) | +336.08 ± 0.20 | −341.46 ± 0.17 | 133.65 ± 0.83 | 1.41 | 0.98 |
| WISE 1828+2650 | 277.1313553(4.0) | +26.84410593(3.3) | +1019.26 ± 1.07 | +172.68 ± 1.00 | 120.00 ± 5.94 | 86.82 | 14.37 |
| WISE 2056+1459 | 314.1217637(0.6) | +14.99896075(0.6) | +825.14 ± 0.12 | +529.23 ± 0.12 | 146.44 ± 0.43 | 1.21 | 1.12 |
| WISE 2209+2711 | 332.2760289(6.6) | +27.19341518(5.8) | +1204.40 ± 2.14 | −1362.08 ± 1.34 | 165.09 ± 12.01 | 10.96 | 1.28 |
| WISE 2220−3628 | 335.2310765(1.2) | −36.47168862(0.8) | +288.92 ± 0.15 | −97.10 ± 0.16 | 93.50 ± 1.02 | 1.29 | 1.18 |

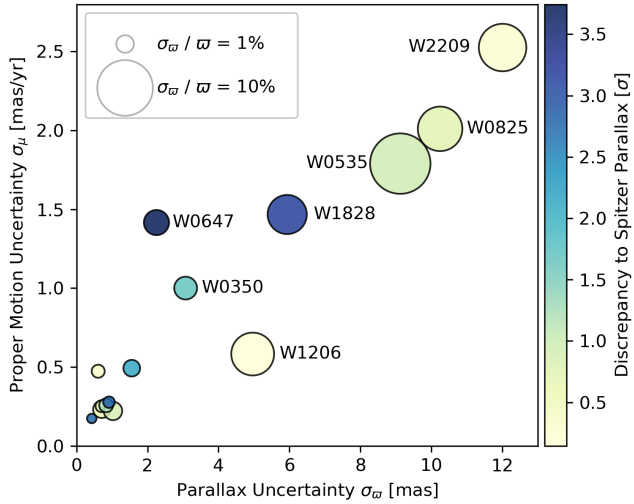

**Figure 2.** Uncertainties in the parallaxes and proper motions for the astrometric solutions derived in this work. The sizes of the markers are proportional to the relative uncertainties in the obtained parallaxes, and the colours indicate the level of consistency between our *HST* results on the targets' parallaxes and those derived from *Spitzer* in Kirkpatrick et al. (2021), based on a quadrature addition of both measurements. All targets with final parallax uncertainties >2 mas are labelled.

Our final astrometric solutions are given in Table 2. We note that the estimated parallaxes are already in an absolute reference system. Figure 1 illustrates the derived solution for WISE 0350−5658, with the full sky motion, with and without the parallax contribution, shown in the left panel, and the proper motion subtracted parallax ellipse in the right panel. Similar plots are shown for all targets in the Figures in Appendix A.

## 4 RESULTS AND DISCUSSION

### 4.1 Overview of Astrometric Results

We present the results from our astrometric fits to all 15 targets in Table 2, in which we list the Y dwarfs positions at epoch 2016.0 (that of *Gaia* DR3), proper motions, and absolute parallaxes. In Figure 2, we show the obtained uncertainties in parallax and proper motion, along with their consistency to the *Spitzer*-derived parallaxes from Kirkpatrick et al. (2021), which we adopt as the nominal reference for comparison with literature astrometry for the Y dwarfs in our program.

Our achieved parallax uncertainties fall below 1–2 mas for 8 of the 15 targets, corresponding to relative precisions of <1% reached for more than half of the sample, shown in the bottom left corner of Figure 2. All of these are in good agreement with the *Spitzer* results, within 0.3–2$\sigma$ for 6 of these 8 systems, and consistent at the 2.8-$\sigma$ level for the remaining 2 objects (WISE 1639−6847 and WISE 2056+1459). For all these, our achieved parallax uncertainties represent 1.4–5× improvements in precision over previous literature measurements (Kirkpatrick et al. 2019, 2021) (see also the Y dwarf Compendium). Our obtained proper motions for these targets have total uncertainties (combining the components in quadrature) of 0.2–0.5 mas/yr, and are all consistent with the *Spitzer*-derived values from

Kirkpatrick et al. (2021) to less than 1–3 mas in each component, or offsets <0.5% in most cases.

The remaining 7 targets, for which we achieved lower precision levels, are all identified by name in Figure 2. We reached slightly larger uncertainties in parallax (∼2–3 mas, 1.7–2% relative uncertainty) and proper motion (0.7–1.2 mas/yr in each component) for WISE 0647−6232 and WISE 0350−5658, consistent at 3.7 and 1.7$\sigma$ with the parallaxes values from Kirkpatrick et al. (2021), respectively, using a quadrature addition of the uncertainties from the *HST* and *Spitzer* estimates. The obtained proper motions for these two targets are in good agreement with those from *Spitzer*, with disparities <1–3 mas/yr, similar to the targets in the best half of the sample, discussed above.

The last 5 targets (WISE 1206+8401, WISE 1828+2650, WISE 0535−7500, WISE 0825+2805 and WISE 2209+2711) have poorer results, with parallax uncertainties ranging from 5–12 mas, corresponding to relative precisions of 5–12% as illustrated by the significantly larger scatter points in Figure 2. With such larger uncertainties, our obtained values are in good agreement with those from Kirkpatrick et al. (2021), to <1-$\sigma$, except for WISE 1828+2650 with a derived parallax 3.1-$\sigma$ away from the *Spitzer* results, with a significant 20-mas (16%) offset. Proper motion uncertainties fall in the range 1–2 mas/yr in each component for all of these targets, apart from WISE 1206+8401 for which we achieved levels of <0.5-mas/yr. As for the rest of the sample, our derived proper motions for these targets are within 1–3 mas from those estimated in the *Spitzer* work by Kirkpatrick et al. (2021).

We also report in Table 2 reduced $\chi^2$ values for the best fit solutions ($\chi^2_\nu$), using the positional scatter in $X$ and $Y$ within each dataset to normalise the sum of the squared residuals between the observed and expected $(X, Y)$ positions, divided by the number of degrees freedom $\nu$ (2× total number of images - 5 fitted parameters). We also computed weighted reduced $\chi^2$, using the quality-of-fit $Q$ values (as $1/Q^2$), to provide an assessment of our results that is more representative of the true picture from our fits ($\chi^2_{\nu,Q}$). As expected, taking into account the weights from the $Q$ values reduces the $\chi^2$, which are otherwise driven to higher values by the inclusion of poorer measurements that may not be reliable. In most cases, the difference is small (∼0.2–0.3), as a result of the lack of largely deviant measurements in the available datasets. Larger differences are only seen for the few targets with some highly compromised individual measurements with large $Q$ values (e.g., WISE 2209+2711, see Fig. A13), and the $\chi^2_{\nu,Q}$ around 1 for these confirms that the inclusion of the weights in our fits leads to more reliable results. With the exception of WISE 1828+2650 (see Fig. A11), our weighted reduced $\chi^2$ are all close to 1 (median of 1.09 with a standard deviation of 0.26), confirming the validity of our approach for estimating the astrometric uncertainties propagated onto our final solutions. WISE 1828+2650 is our only poor fit, and is discussed further in Section 4.4.

In the following Sections, we explore various effects to identify and assess the dominant factors impacting the results described above, and discuss individual targets.

### 4.2 Assessment of Performance

#### 4.2.1 Effect of Signal-to-Noise Ratio

Given the faintness of our observed Y dwarfs in the near-IR, even with deep *HST* observations, the signal-to-noise ratios (SNRs) achieved across the various available datasets for each object is likely to be one of the main limiting factors for our astrometric results, with lower precisions in individual positional measurements expected for low

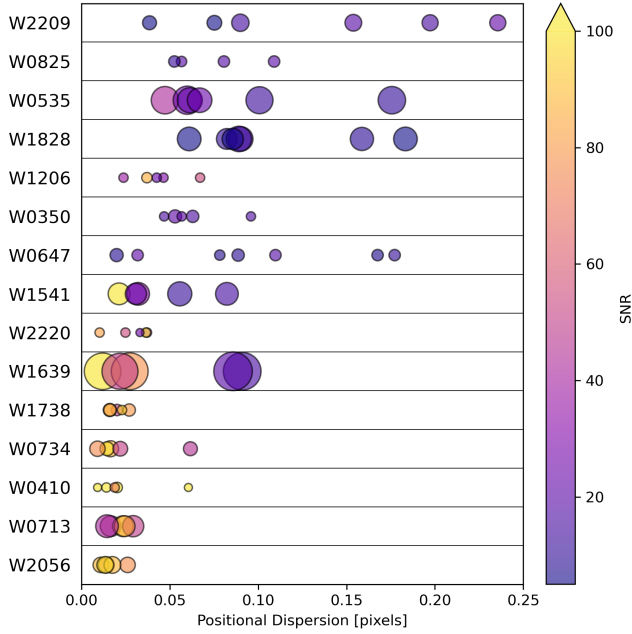

**Figure 3.** Estimates of the positional dispersion within each *HST* dataset for each target, ordered from highest (top) to lowest (bottom) final uncertainty in the derived parallaxes. Each scatter point corresponds to the images from a given epoch and filter combination, as listed in Table 1, with the colour bar indicating the average signal-to-noise ratio of the Y dwarf in the considered dataset. The sizes of the symbols are proportional to the minimum number of Gaia stars available among the images from any given dataset. Multiple epochs with high SNRs (orange-yellow colours), strongly correlated with lower positional dispersions, were found to be needed to reach high precisions in the derived astrometric solutions (bottom rows).

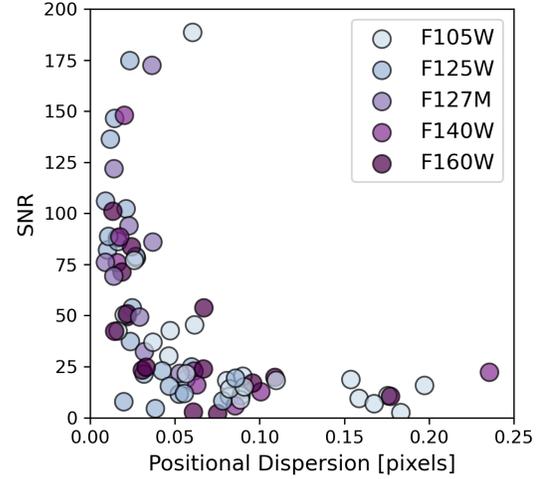

**Figure 4.** Positional uncertainties within each dataset as a function of the average target SNR in the corresponding images, colour coded by the filter of the considered observations.

SNR datasets. Another restricting component in our analyses may be the number of *Gaia* stars available in the images as astrometric references to link the *HST* and *Gaia* frames. Finally, the number of epochs, yearly phases and time baseline covered by the available data may also be impacting the final precisions that can be reached in the derived astrometric solutions.

In Figure 3, we test the effects of target SNRs, along with dependence on the number of reference stars. Each row corresponds to a single target from our sample, ordered from largest to smallest parallax uncertainty reached in our final solutions. For each epoch, all images in a given filter (see Table 1) are grouped into a single dataset, as we did in our uncertainty derivation from Section 3.2 (but the specific dates of each observation are considered in the astrometric analysis). The average SNR from that dataset is then estimated from the standard deviation in the instrumental magnitude of the target in each image (SNR $\sim 1/\sigma_{\rm mag}$). We estimate the positional dispersion within that dataset from the standard deviation in the measured X and Y positions of the target in each exposure within that epoch and filter. Each individual dataset is plotted as a separate circle in Figure 3, where its position along the x-axis corresponds to that positional dispersion (in pixels), and the average SNR is shown with the colour bar, ranging from SNR~5 in the worst cases (purple) up to >100 for the brightest targets and deepest observations (yellow). Finally, the size of each scatter point is proportional to the minimum number of *Gaia* reference stars available among the images from that dataset, ranging from only 3 for WISE 0410+1502 and WISE 2220−3628 to 60 for WISE 1639−6847. Similar numbers of *Gaia* stars are typically available across the various datasets for each target, depicted by the similar symbol sizes within each row.

We find that the half of the sample for which we reached better than 2 mas on the final parallaxes (i.e., the lower half of the Figure, from WISE 1541−2250 to WISE 2056−1459) all have at least 3 or more datasets with positional dispersions <0.04 pixels (<5 mas). In fact, further exploring the individual epochs from these datasets, we found that all of these targets have at least 3 *distinct epochs* with one or more dataset achieving this positional uncertainty level. As expected, these robust positional measurements are well correlated with the high SNRs (>40–100) achieved in these datasets, as seen from the pink-orange-yellow colours of all these points in Figure 3. The high precision reached in the final astrometric solutions for these objects shows little dependence on the number of reference stars available, with several of these targets having some of the lowest numbers of available *Gaia* stars out of the full sample. We therefore conclude that having at least 3 epochs with high SNRs is a primary requirement to achieve the highest precision levels in the final astrometric solutions with our methods. This is consistent with predictions from King (1983) that SNR is the driving factor as long as other sources of noise are less important.

The next 3 targets (WISE 0647−6232, WISE 0350−5658 and WISE 1206+8401) have slightly higher final parallax precisions of 2–5 mas, which we attribute to the lower SNRs of these objects (<10–20 in most cases), resulting in larger positional dispersions above 0.05 pixels for most of the individual datasets and up to 0.1–0.2 pixels (12–24 mas), in addition to low numbers of available *Gaia* reference stars in the range 4–8. The remaining targets, at the top of Figure 3 have very low SNRs (purple symbols) and positional offsets >0.05–0.2 for almost all available datasets, resulting in poorer final astrometric precisions of 6–12 mas in parallax, despite significantly higher numbers of astrometric reference sources for WISE 0535−7500 and WISE 1828+2650 (>20–30 *Gaia* stars).

It should also be noted that the majority of the archival programs used here did not collect images with astrometry in mind, and the lack of robust dithering in the observing strategy could compromise the accuracy of the obtained astrometry for case with only 3 epochs, even if high SNRs are available. In Figure 4, we show the measured positional uncertainties within each epoch and filter as a function of the average SNR of our targets in the considered images. Positional uncertainties range from 0.02–0.25 pix (2–30 mas) at SNRs below ~25, and drop to scatters <0.06 pix (7 mas) for SNRs~25–50, and down to a systematic floor of 0.01–0.04 pix (~1–5 mas) for SNRs>50. These measurements are in good agreement with previous investigations of WFC3/IR astrometric uncertainties as a function of SNR or source brightness (Anderson 2016). We also note no trend with filters, highlighting that any wavelength-dependent variations must represent a minor contribution towards the overall uncertainties, as long as similar SNRs are reached between bands.

#### 4.2.2 Effect of Epoch Sampling

In Figure 5, we explore the dependence of our results on the number of main observational epochs available and coverage of the parallax ellipses and proper motion displacements. With our *HST* program (GO 16229), we were able in most cases to schedule a new epoch at the same time of year as one of the previous archival datasets, optimally sampling the proper motion of the targets. This is illustrated in the left panel of Figure 5, where most targets have <10–15 days between the yearly phases of two separate epochs from different years, and <5 days for more than half of the sample. Similarly, we achieved an ideal sampling of the parallax motion through datasets obtained at opposite times of year for the majority of the sample, with these epochs probing the maximum elongation of the parallax ellipses within ~2 weeks in most cases.

The only target for which we are limited to a less optimal sampling of the astrometric displacements is WISE 0825+2805, with 1 month separating the closest times of year probed, and 4 months between the longest yearly sampling. This target is our second poorest in terms of final parallax uncertainty, but is also the only object for which we are limited to the bare minimum of 3 observational epochs, and suffered from low SNRs and small numbers of available *Gaia* stars. Looking at the rest of the targets, no strong correlation appears between the temporal baselines of the available epochs, and the final precisions achieved in our astrometric fits (indicated by the size of the scatter points in Figure 5, which are the same as in Figure 2). We therefore conclude that within the ranges of temporal samplings available in the rest of our sample, the number of epochs (from 4 to 6), total time baseline (7–10 years), and exact timings of observations, do not appear to be the dominant factors impacting the astrometric precisions we were able to reach.

### 4.3 Unresolved multiplicity

Finally, unknown multiplicity for our science targets could also lead to spurious astrometric solutions. Late-type binaries in the field show a strong tendency towards tight orbital separations, with an observed peak around ~3 au for mid-to-late T and Y dwarfs (Fontanive et al. 2018). Recently, the first companion around a Y-type primary was discovered with *JWST* (Calissendorff et al. 2023), with a projected separation just under 1 au in agreement with this trend. While this system, WISE 0336−0143 AB, is unresolved in *HST* WFC3/IR observations due to its tight 84-mas angular separation and the faintness of the cold secondary in the near-IR, the companion is expected to induce a ~30-mas offset on the observed absolute astrometry of the photocentre due to from wobble of the binary orbit. This represents about a third of the parallax displacement for a system located at 10 pc, which would likely lead to an erroneous derivation of the astrometric parameters without taking the binary nature of the system into account and with only few observational epochs as done in this study. For this reason, we left WISE 0336−0143 out from this work, despite it being part of our original sample, and will require additional observational epochs to simultaneously fit the absolute astrometry of the system and the binary orbit (GO 17466).

While none of the other Y dwarfs from our program are currently known to be binaries, the discovery of WISE 0336−0143 highlighted the need to probe the very tightest separations to uncover Y-dwarf binaries. It is hence possible that more close systems currently unresolved still lie within the known Y dwarf population. A few of our targets have been noted in the literature to be possible binaries, which we list in the next Section. Our poorer astrometric fits to some of these could be a further indication of the possible binary nature of some of these systems, including WISE 0535−7500 and WISE 1828+2650.

### 4.4 Notes on individual targets

WISE 0535−7500: this target passed close to a background source during 2 of the main 5 available *HST* epochs, which were providing the ideal proper motion sampling needed for optimal results, leading to degraded measurements (with average $Q \sim 0.2$–0.5 within these datasets), and a poorer astrometric solution despite >30 *Gaia* reference stars available in most images. This object was noted by Tinney et al. (2014) and Kirkpatrick et al. (2019, 2021) to be overluminous and a potential binary (Leggett et al. 2017), although equal-mass companions >1.9 au were ruled out by Opitz et al. (2016).

WISE 0825+2805: this object has a low SNR in our F160W images (dataset mean of $Q \sim 0.3$), resulting in a large dispersion in positions, consistent with the conclusions from Kirkpatrick et al. (2019, 2021) that this Y dwarf may be underluminous in *H*-band for it spectral type. Our final astrometric precision for this object is also likely affected by the fact that we only have 3 main observational epochs, since this target already had 2 available epochs in 2 of our 3 filters of interest (F105W and F125W), and only one final epoch in F160W was requested in our program according to the strategy for our campaign design. Unfortunately, due to the timings of the archival epochs, the final 3 epochs available do not provide a complete sampling of the parallax ellipse nor proper motion displacement. This is added to the fact that it has the second shortest time baseline of all targets and the lowest number of total images available, in addition to rather low numbers of reference stars in the field of view, varying from 5–8 per image.

WISE 1828+2650: this object is among our lowest-SNR targets, with low-quality PSF fits in most images (dataset mean of $Q \sim 0.4$–0.8) and large resulting positional scatters within individual epochs. This target has the worse weighted reduced $\chi^2$ of our sample ($\chi^2_{\nu,Q} \sim 14.37$), and is the only one with $\chi^2_{\nu,Q} > 2$. The photometric and spectroscopic characterisation of this Y dwarf has been particularly problematic, with the near impossibility of simultaneously fitting both near- and mid-infrared observations of this object (Kirkpatrick et al. 2012; Dupuy & Kraus 2013; Kirkpatrick et al. 2019). It has been suggested to be a suspected equal-magnitude binary (Beichman et al. 2013; Leggett et al. 2013; Cushing et al. 2021), although De Furio et al. (2023) ruled out companions beyond 0.5 au. Unresolved binarity could impact accurate centering of the PSF and potentially explain some of the very large positional dispersion seen within each HST epoch (see Fig. A11). We also note that only the latest observational epoch, from our program, is optimised for astrometry, and most of the remaining archival images were obtained for spectroscopy and were thus not designed with astrometry in mind, therefore leading to particularly large errors, especially given the low SNR of this target at *HST* wavelengths.

WISE 2209+2711: this Y dwarf is among the faintest of our sample in *H* band (Kirkpatrick et al. 2021), resulting in low SNRs in the rather shallow exposures from archival observations. In addition, the target unfortunately passed close to the diffraction spike of bright star in the latest epoch, from our program. While we had identified that the target's sky trajectory was approaching a bright object in the field, and scheduled the observations as early as possible in the *HST* cycle to mitigate this issue, this still resulted in a somewhat degraded quality for the datasets of the latest epoch ($Q \sim 0.6$).

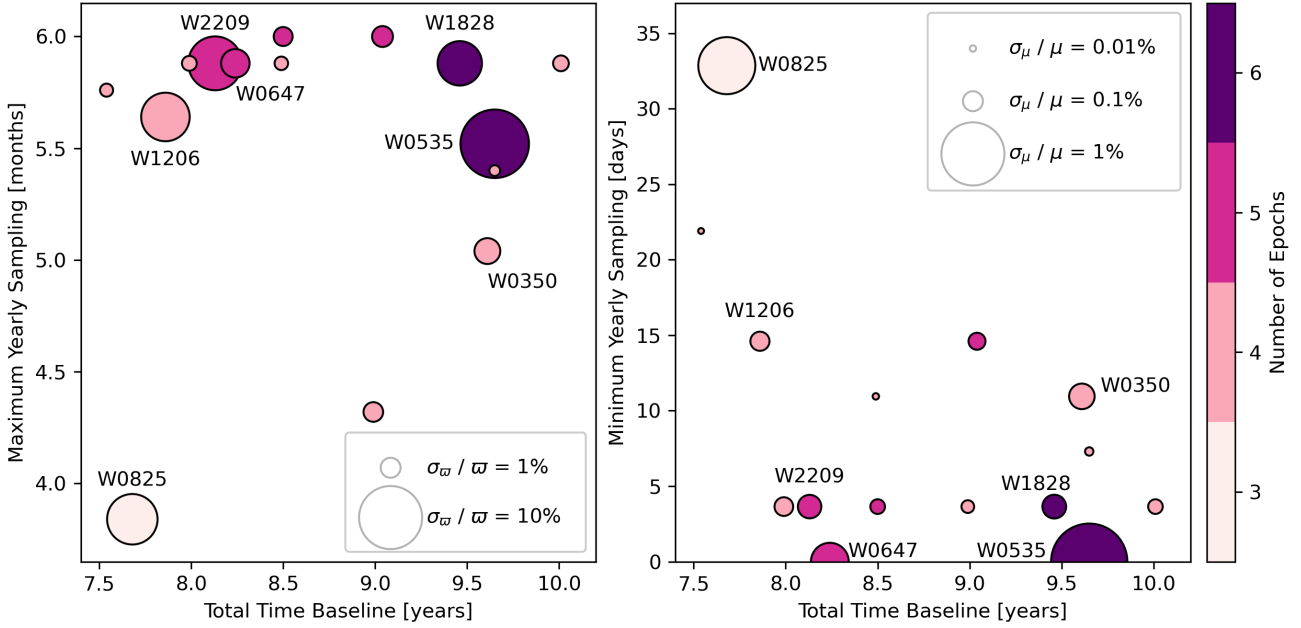

**Figure 5.** Total time baseline covered by the *HST* datasets considered here, plotted against the maximal (left) and minimal (right) intervals between the yearly phases sampled by the available epochs, i.e., the best samplings of the parallax ellipse and proper motion, respectively. For the minimum yearly sampling, only datasets from different years are considered. Symbol sizes are proportional to the relative parallax (left) and proper motion (right) uncertainties. The discrete colour bar represents the number of main observational epochs available, as given by the individual lines in Table 1.

## 5 CONCLUSIONS

In this work, we used multi-epoch *HST* observations to derive accurate astrometric solutions for 15 nearby Y-type brown dwarfs, by anchoring the relative *HST* astrometry to the *Gaia* DR3 absolute reference system, providing some of the most precise distances to date for 8 new Y dwarfs.

Our parallax and proper motion estimates generally align well with recent literature values, typically within 1–2 $\sigma$ of *Spitzer*-derived parallaxes by Kirkpatrick et al. (2021). Our powerful approach achieved parallax uncertainties below 1 milliarcsecond for over half of the sample, and down to 3 mas for two thirds of the targets. This translates to relative precisions of less than 1%, marking a substantial improvement over previous measurements, by a factor of 2–5× for these objects. However, some targets showed lower precision, with parallax uncertainties ranging from 5 to 12 mas (5–10%). Our derived proper motions typically showed total uncertainties of 0.2–0.5 mas/yr, consistent with *Spitzer*-derived values within 1–3 mas/yr in all cases.

We attribute the performance variations among our sample to the signal-to-noise ratios of the science targets in the available observations as the dominant effect, with the availability of high numbers of *Gaia* reference stars, and temporal coverage of the observational epochs as secondary limiting factors. We found that achieving high precision (better than 2 mas) in parallax measurements consistently required at least 3 distinct epochs with high SNRs (above 40). Temporal spacing of observational epochs, while generally well-sampled across our program, did not strongly correlate with final astrometric precisions. Finally, unresolved multiplicity could be affecting our derived astrometric solutions, and some of our poorer astrometric fits may be a further indication of binarity for some suspected binaries in our sample.

Our study underscores the effectiveness of *HST* in conjunction with *Gaia* to obtain precise astrometry for faint Y-type brown dwarfs. These new distance measurements will be fundamental to explore the absolute fluxes of these objects and peak into the structure and dynamics of their atmospheres, which will be the focus of the next paper in this series of *HST* work. Our results also highlight the challenges encountered with the reddest and coldest objects with *HST*. The growing pool of Y-dwarf observations with *JWST*, with an enhanced sensitivity to such cold systems, will offer new opportunities to refine these measurements and further advance our understanding of these intriguing objects, especially for the coldest and faintest targets. This is particularly important at the dawn of a new era of exoplanet studies, where we can finally directly image Y-temperature giant exoplanets on wide orbits around stars, as recently demonstrated with the *JWST*/MIRI direct detection of the 275-K super-Jupiter around Eps Ind A by Matthews et al. (2024).

**DATA AVAILABILITY**

All observational data used in this work are publicly available. The reduced stacked images produced for each epoch/filter combination for each target are provided at https://github.com/cfontanive/BDs_HST.

10     *C. Fontanive et al.*## ACKNOWLEDGEMENTS

We are grateful to the referee for their careful review of our work that contributed to improve our manuscript. CF acknowledges support from the Trottier Family Foundation and the Trottier Institute for Research on Exoplanets through her Trottier Postdoctoral Fellowship. The authors acknowledge the support from STScI award #HST-GO-16229.002-A. This research is based on observations made with the NASA/ESA *Hubble Space Telescope* obtained from the Space Telescope Science Institute, which is operated by the Association of Universities for Research in Astronomy, Inc., under NASA contract NAS 5–26555. These observations are associated with programs GO 16229, 15201, 13178, 12970, 12873, 12544, 12330. Some of the data presented in this paper were obtained from the Mikulski Archive for Space Telescopes (MAST). This work has made use of data from the European Space Agency (ESA) mission *Gaia* (https://www.cosmos.esa.int/gaia), processed by the *Gaia* Data Processing and Analysis Consortium (DPAC, https://www.cosmos.esa.int/web/gaia/dpac/consortium). Funding for the DPAC has been provided by national institutions, in particular the institutions participating in the *Gaia* Multilateral Agreement. This research has benefited from the Y Dwarf Compendium maintained by Michael Cushing at https://sites.google.com/view/ydwarfcompendium/.
## REFERENCES

Anderson J., 2016, Empirical Models for the WFC3/IR PSF, Space Telescope WFC Instrument Science Report
Anderson J., King I. R., 2006, Technical report, PSFs, Photometry, and Astronomy for the ACS/WFC
Anderson J., et al., 2008, AJ, 135, 2055
Bardalez Gagliuffi D. C., et al., 2020, ApJ, 895, 145
Bedin L. R., Fontanive C., 2018, MNRAS, 481, 5339
Bedin L. R., Fontanive C., 2020, MNRAS, 494, 2068
Beichman C., Gelino C. R., Kirkpatrick J. D., Barman T. S., Marsh K. A., Cushing M. C., Wright E. L., 2013, ApJ, 764, 101
Beichman C., Gelino C. R., Kirkpatrick J. D., Cushing M. C., Dodson-Robinson S., Marley M. S., Morley C. V., Wright E. L., 2014, ApJ, 783, 68
Beiler S. A., Cushing M. C., Kirkpatrick J. D., Schneider A. C., Mukherjee S., Marley M. S., Marocco F., Smart R. L., 2024, arXiv e-prints, p. arXiv:2407.08518
Calissendorff P., et al., 2023, ApJ, 947, L30
Cushing M. C., et al., 2011, ApJ, 743, 50
Cushing M. C., Kirkpatrick J. D., Gelino C. R., Mace G. N., Skrutskie M. F., Gould A., 2014, AJ, 147, 113
Cushing M. C., et al., 2021, ApJ, 920, 20
De Furio M., et al., 2023, ApJ, 948, 92
Dupuy T. J., Kraus A. L., 2013, Science, 341, 1492
Dupuy T. J., Liu M. C., Leggett S. K., 2015, ApJ, 803, 102
Eisenhardt P. R. M., et al., 2020, ApJS, 247, 69
Fontanive C., Biller B., Bonavita M., Allers K., 2018, MNRAS, 479, 2702
Fontanive C., Bedin L. R., Bardalez Gagliuffi D. C., 2021, MNRAS, 501, 911
Fontanive C., Bedin L. R., De Furio M., Biller B., Anderson J., Bonavita M., Allers K., Pantoja B., 2023, MNRAS, 526, 1783
Gaia Collaboration et al., 2016, A&A, 595, A1
Gaia Collaboration et al., 2018, A&A, 616, A1
Gaia Collaboration et al., 2023, A&A, 674, A1
Kaplan G., Bartlett J., Monet A., Bangert J., Puatua W., 2011. User's Guide to NOVAS Version F3.1. USNO, Washington, DC
King I. R., 1983, PASP, 95, 163
Kirkpatrick J. D., et al., 2012, ApJ, 753, 156
Kirkpatrick J. D., Cushing M. C., Gelino C. R., Beichman C. A., Tinney C. G., Faherty J. K., Schneider A., Mace G. N., 2013, ApJ, 776, 128
Kirkpatrick J. D., et al., 2019, ApJS, 240, 19
Kirkpatrick J. D., et al., 2021, ApJS, 253, 7
Lacy B., Burrows A., 2023, ApJ, 950, 8
Leggett S. K., Morley C. V., Marley M. S., Saumon D., Fortney J. J., Visscher C., 2013, ApJ, 763, 130
Leggett S. K., Tremblin P., Esplin T. L., Luhman K. L., Morley C. V., 2017, ApJ, 842, 118
Leggett S. K., et al., 2021, ApJ, 918, 11
Liu M. C., Dupuy T. J., Bowler B. P., Leggett S. K., Best W. M. J., 2012, ApJ, 758, 57
Luhman K. L., 2014, ApJ, 786, L18
Luhman K. L., Esplin T. L., 2016, AJ, 152, 78
Luhman K. L., Burgasser A. J., Bochanski J. J., 2011, ApJ, 730, L9
Luhman K. L., et al., 2024, AJ, 167, 5
Marley M. S., et al., 2021, ApJ, 920, 85
Marocco F., et al., 2019, ApJ, 881, 17
Marocco F., et al., 2021, ApJS, 253, 8
Marsh K. A., Wright E. L., Kirkpatrick J. D., Gelino C. R., Cushing M. C., Griffith R. L., Skrutskie M. F., Eisenhardt P. R., 2013, ApJ, 762, 119
Martin E. C., et al., 2018, ApJ, 867, 109
Matthews E. C., et al., 2024, Nature, 633, 789
Meisner A. M., et al., 2020, ApJ, 889, 74
Moré J. J., Garbow B. S., Hillstrom K. E., 1980. User Guide for MINPACK-1. Argonne National Laboratory Report ANL-80-74, Argonne, IL
Opitz D., Tinney C. G., Faherty J. K., Sweet S., Gelino C. R., Kirkpatrick J. D., 2016, ApJ, 819, 17
Phillips M. W., et al., 2020, A&A, 637, A38
Pinfield D. J., et al., 2014, MNRAS, 444, 1931
Schneider A. C., et al., 2015, ApJ, 804, 92
Schneider A. C., Cushing M. C., Kirkpatrick J. D., Gelino C. R., 2016, ApJ, 823, L35
Tinney C. G., Faherty J. K., Kirkpatrick J. D., Wright E. L., Gelino C. R., Cushing M. C., Griffith R. L., Salter G., 2012, ApJ, 759, 60
Tinney C. G., Faherty J. K., Kirkpatrick J. D., Cushing M., Morley C. V., Wright E. L., 2014, ApJ, 796, 39
Tinney C. G., et al., 2018, ApJS, 236, 28
Wright E. L., et al., 2010, AJ, 140, 1868
## APPENDIX A: FIGURES OF THE DERIVED ASTROMETRIC SOLUTIONS

This paper has been typeset from a TeX/LaTeX file prepared by the author.

MNRAS **000**, 1–10 (2024)

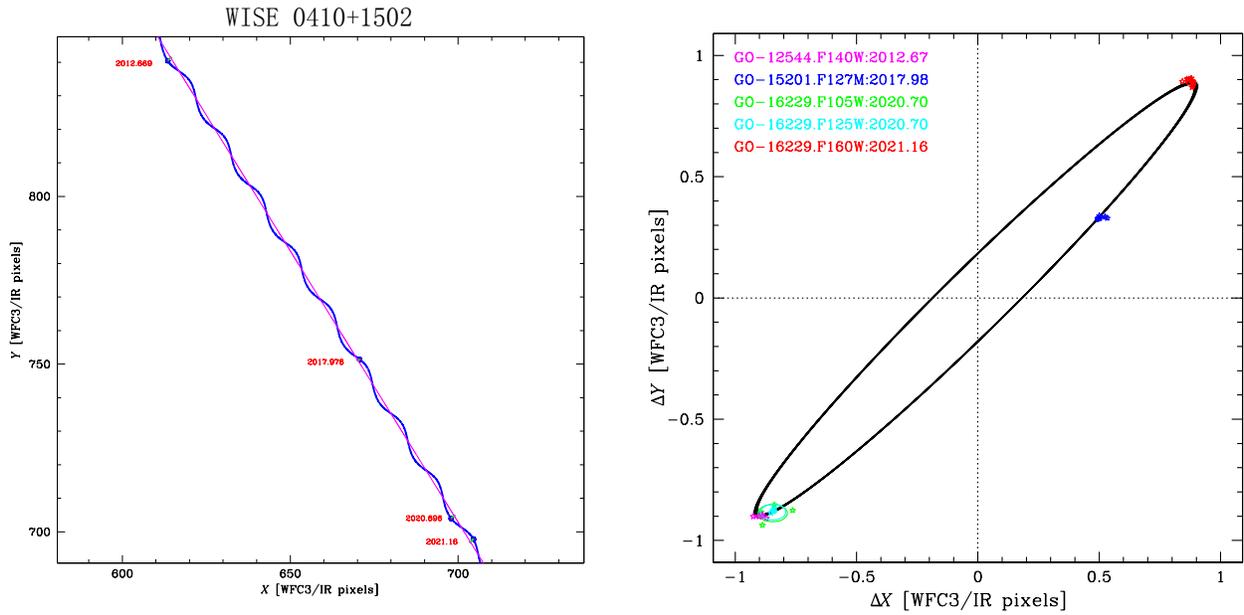

**Figure A1.** Derived astrometric solution for WISE 0410+1502 (same as Figure 1).

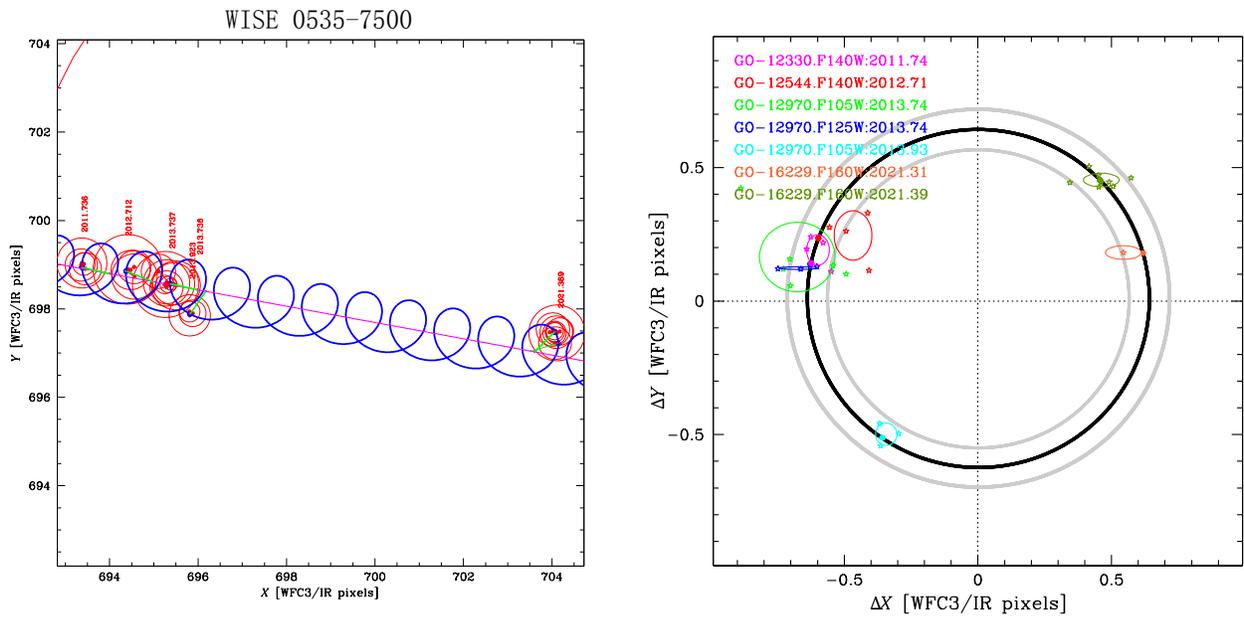

**Figure A2.** Derived astrometric solution for WISE 0535−7500 (same as Figure 1).

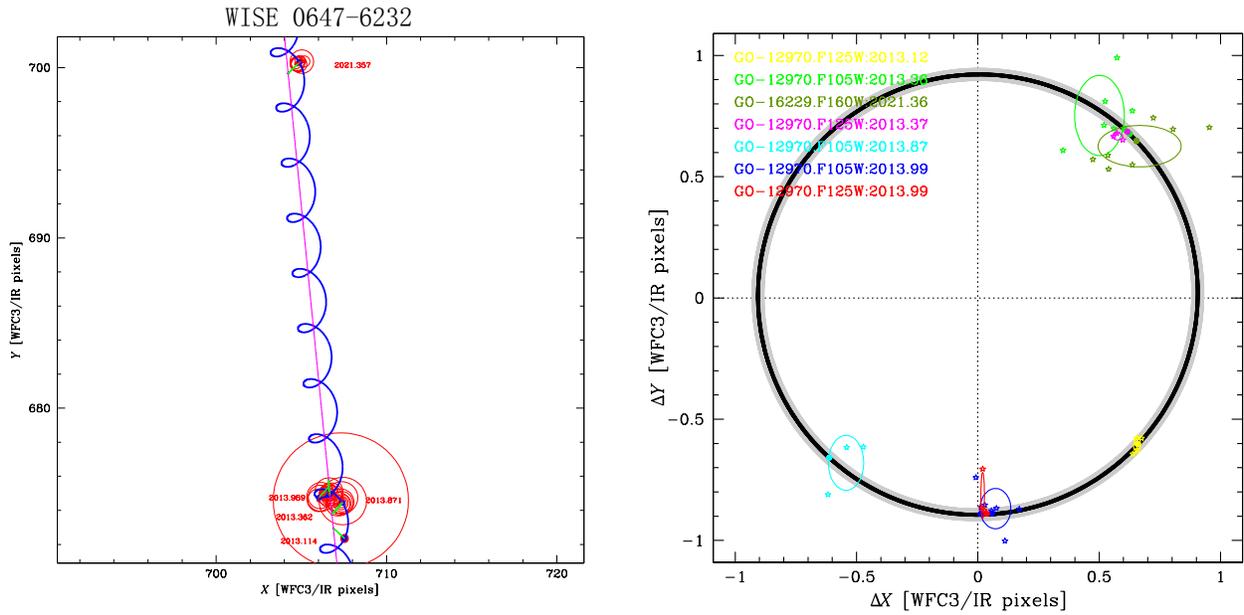

**Figure A3.** Derived astrometric solution for WISE 0647−6232 (same as Figure 1).

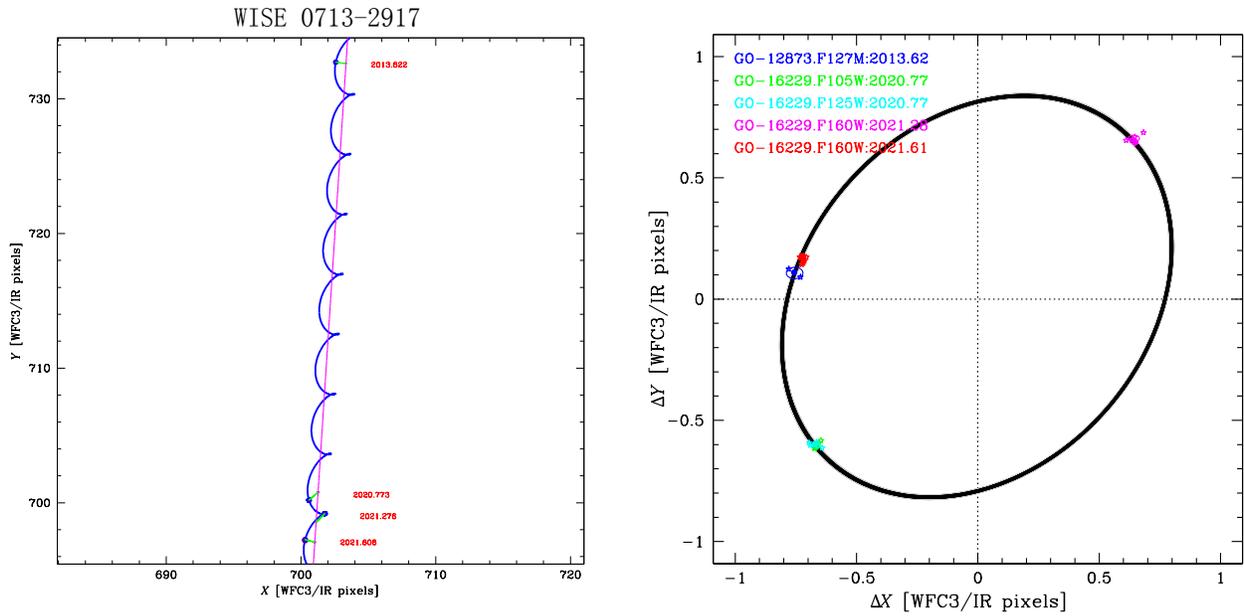

**Figure A4.** Derived astrometric solution for WISE 0713−2917 (same as Figure 1).

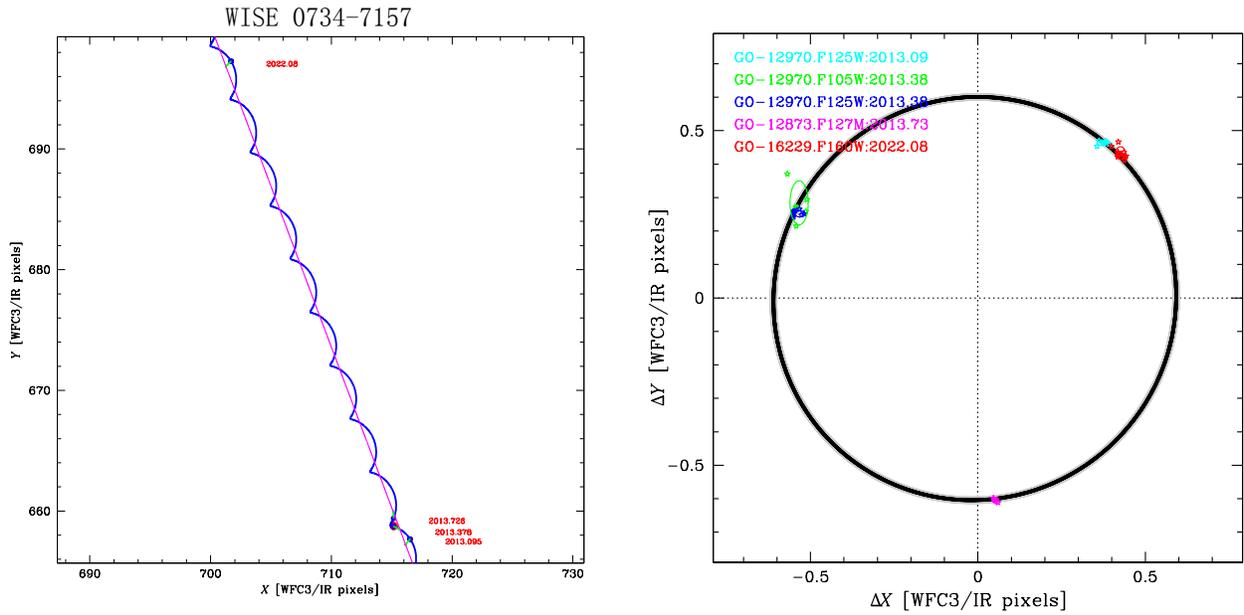

**Figure A5.** Derived astrometric solution for WISE 0734−7157 (same as Figure 1).

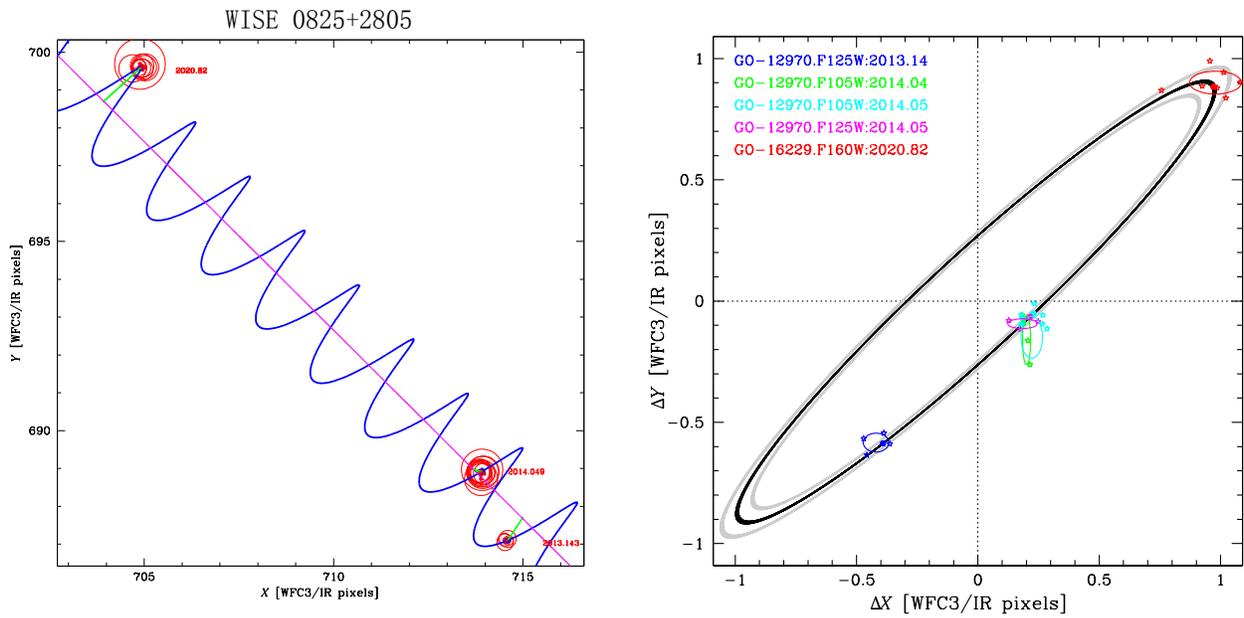

**Figure A6.** Derived astrometric solution for WISE 0825+2805 (same as Figure 1).

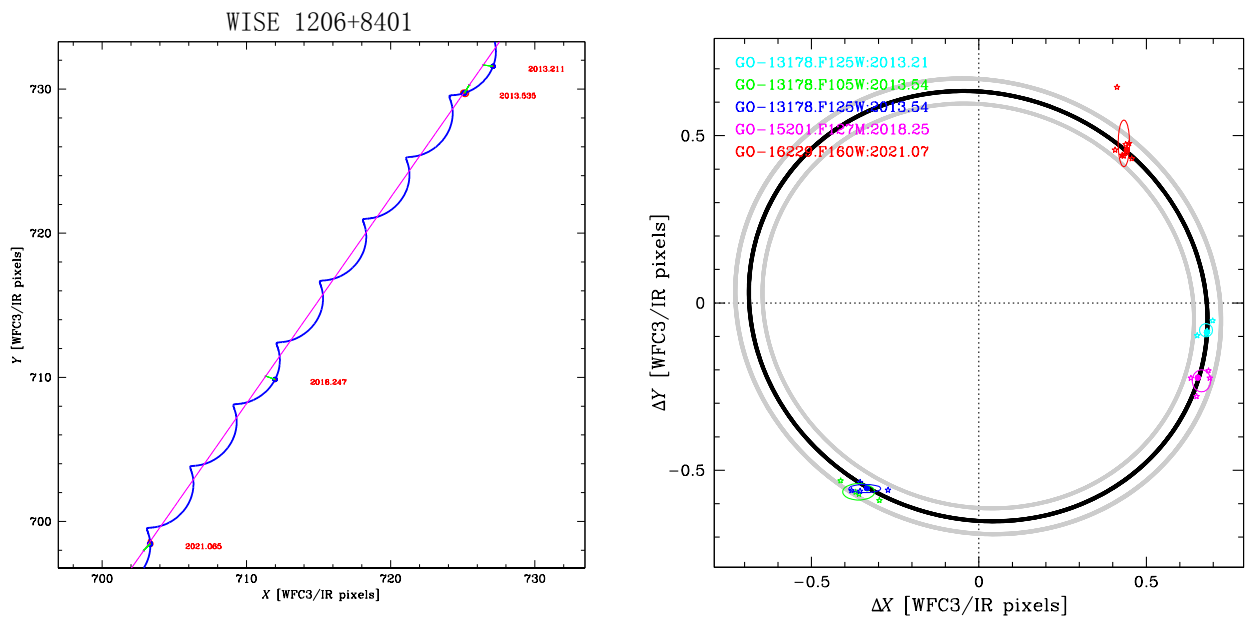

**Figure A7.** Derived astrometric solution for WISE 1206+8401 (same as Figure 1).

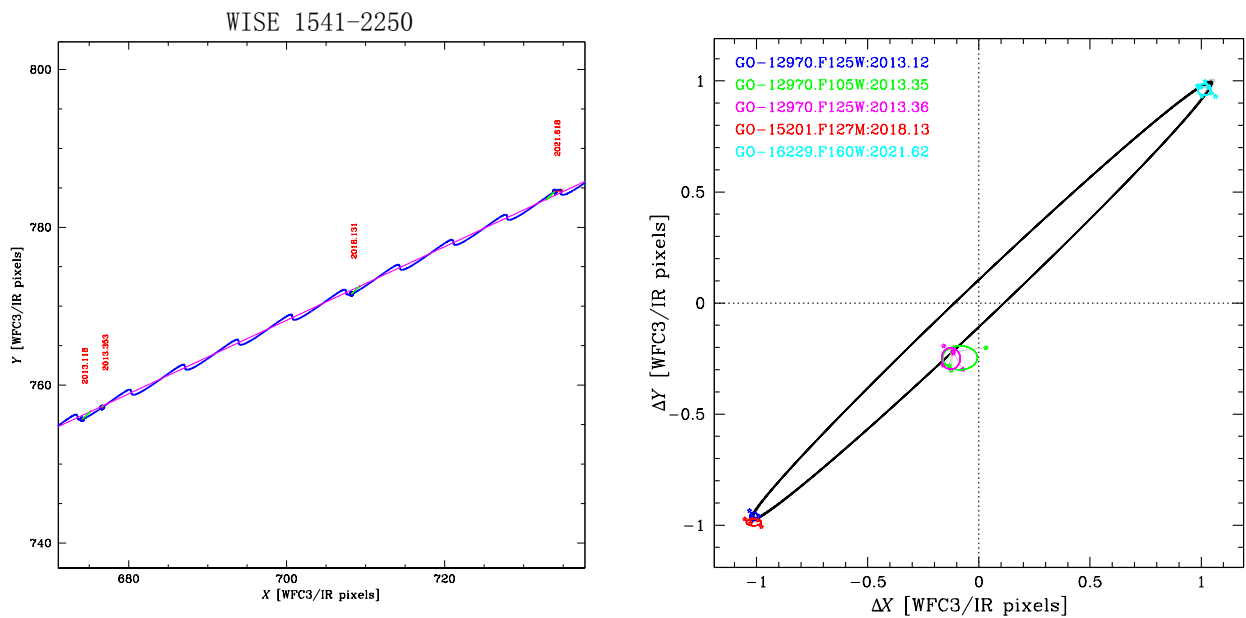

**Figure A8.** Derived astrometric solution for WISE 1541−2250 (same as Figure 1).

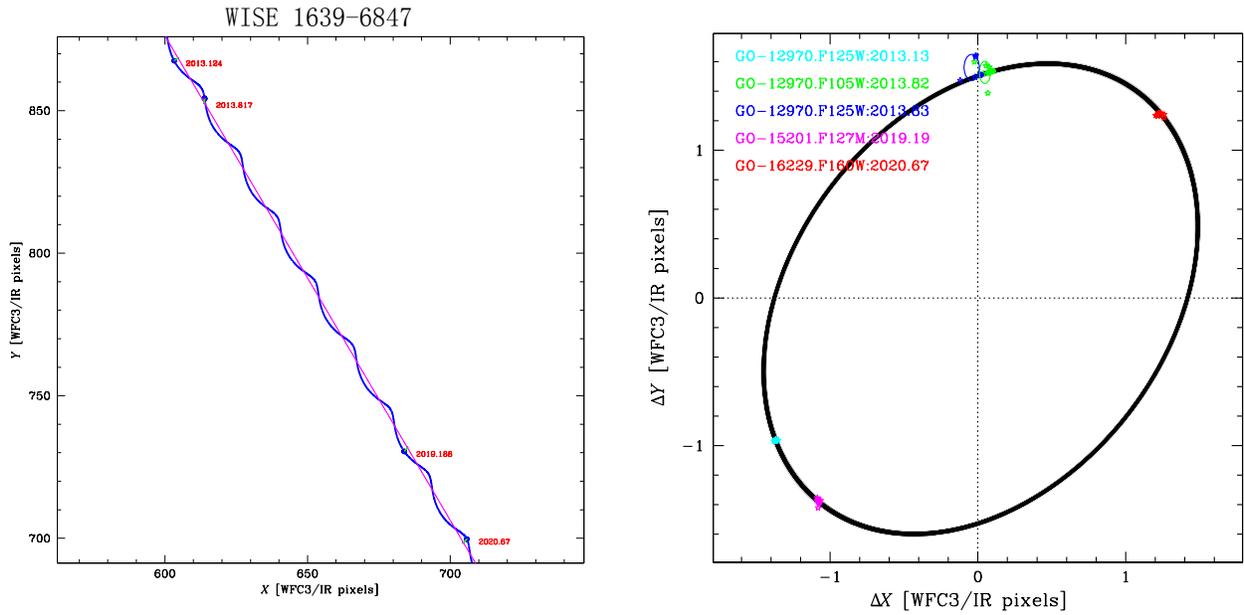

**Figure A9.** Derived astrometric solution for WISE 1639−6847 (same as Figure 1).

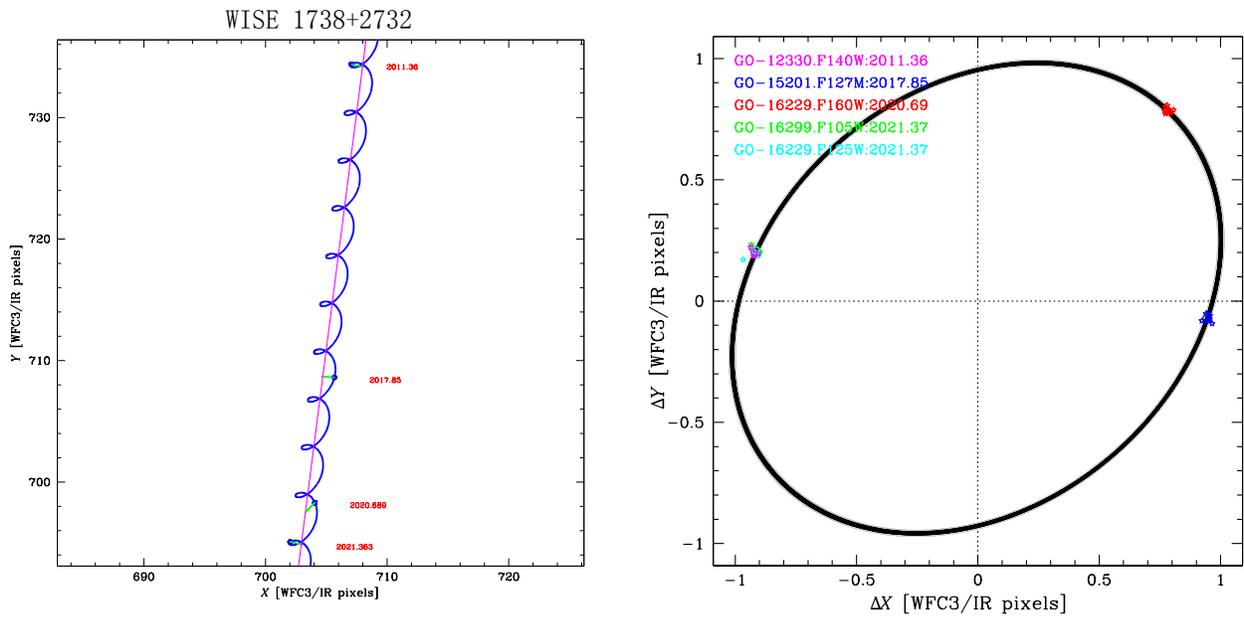

**Figure A10.** Derived astrometric solution for WISE 1738+2732 (same as Figure 1).

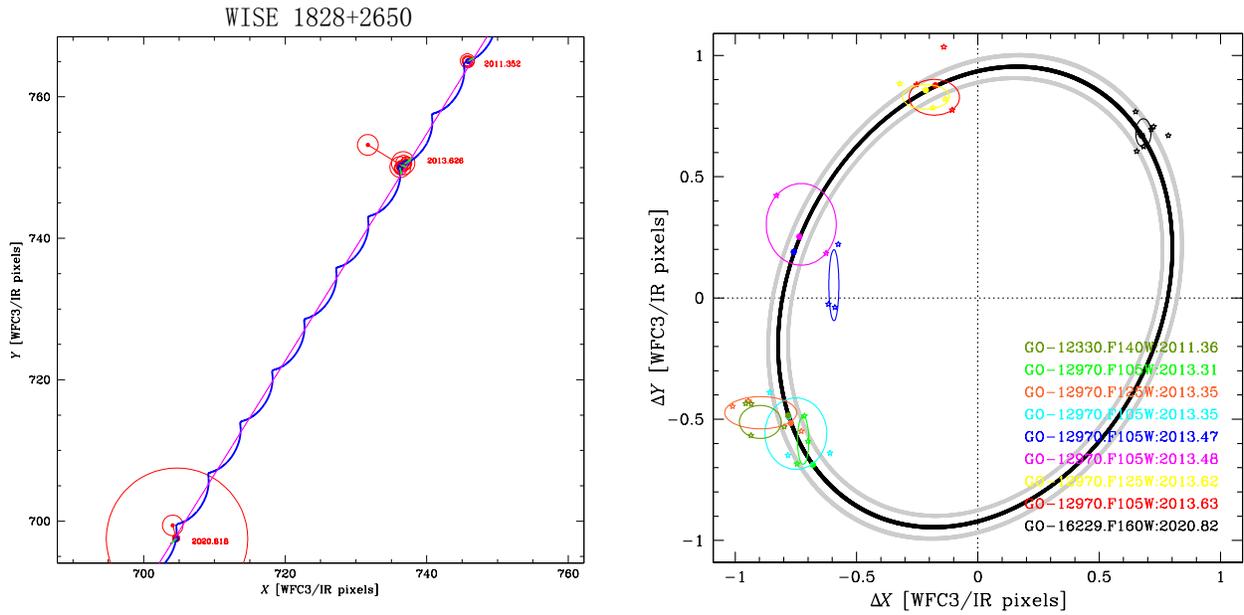

**Figure A11.** Derived astrometric solution for WISE 1828+2650 (same as Figure 1).

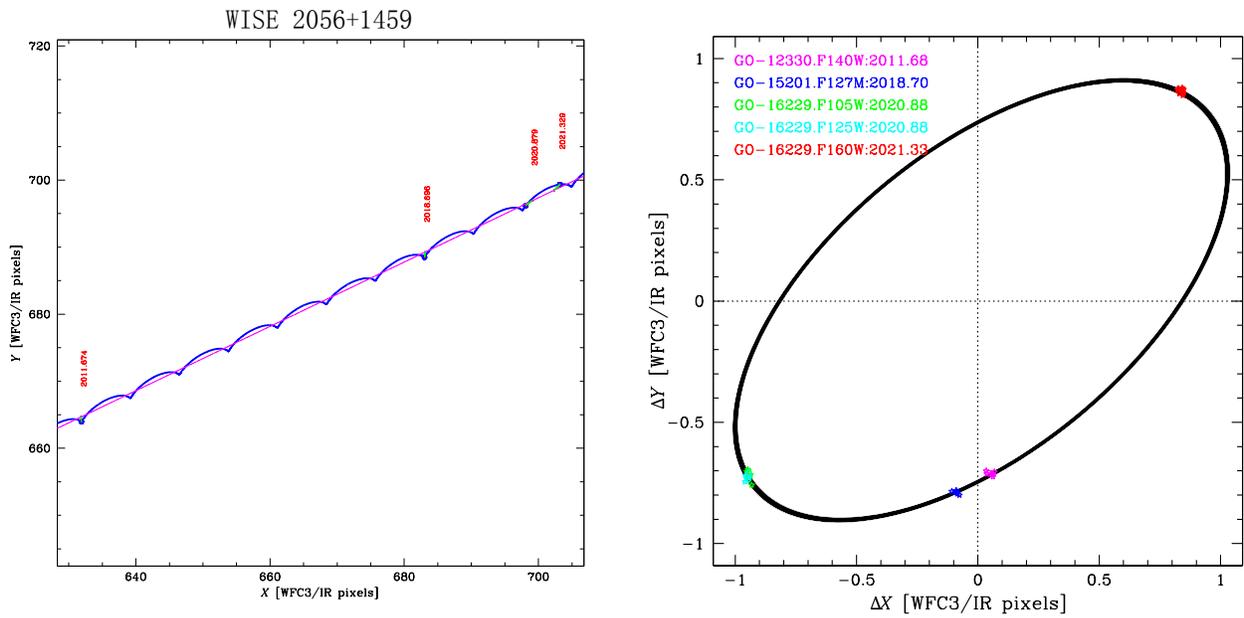

**Figure A12.** Derived astrometric solution for WISE 2056+1459 (same as Figure 1).

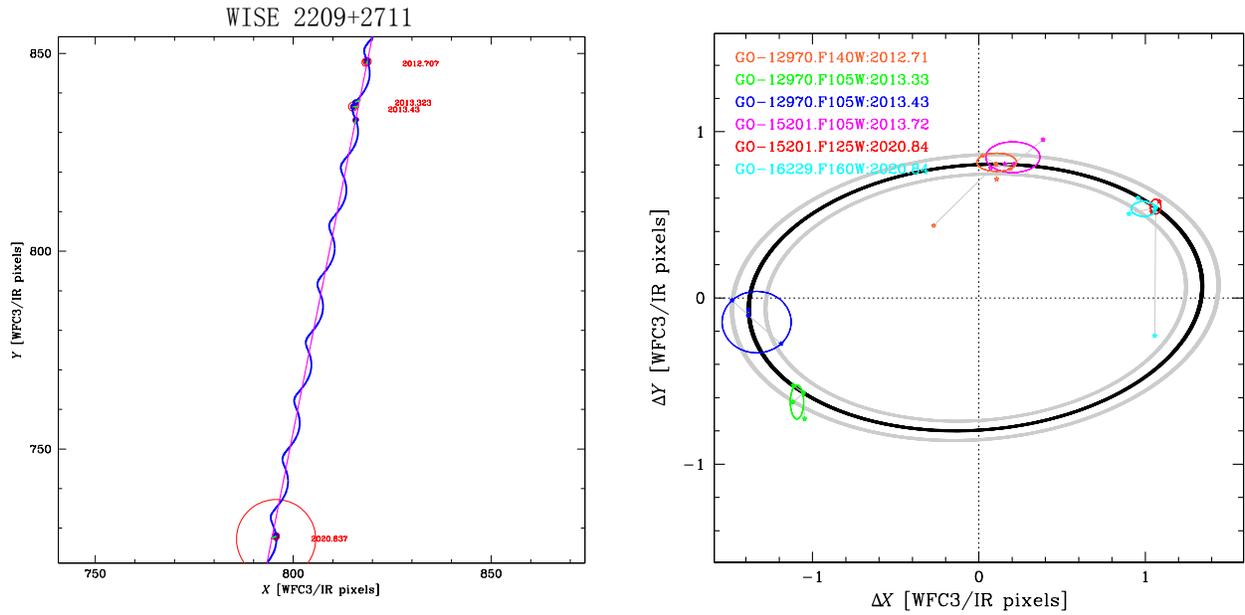

**Figure A13.** Derived astrometric solution for WISE 2209+2711 (same as Figure 1).

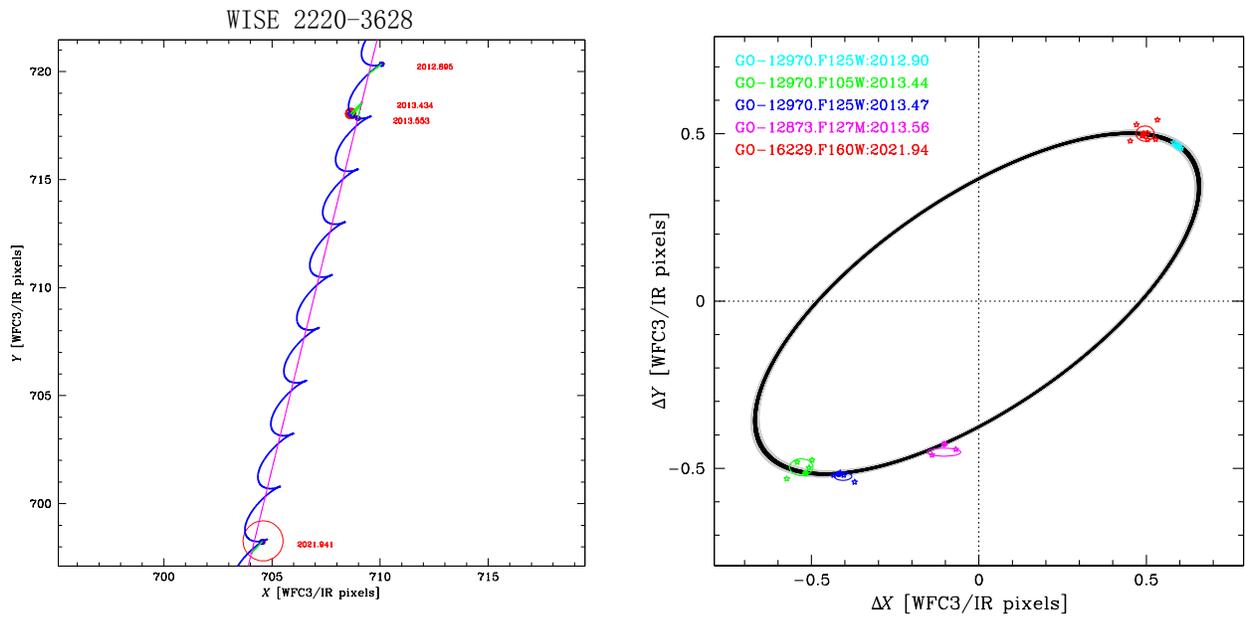

**Figure A14.** Derived astrometric solution for WISE 2220−3628 (same as Figure 1).